\documentclass[aps,prb,twocolumn,graphicx,superscriptaddress]{revtex4-1}

\usepackage{graphicx}
\usepackage{amsmath}
\usepackage{mathtools}

\begin{document}

\title{Optical investigation of thermoelectric topological crystalline insulator Pb$_{0.77}$Sn$_{0.23}$Se}
\author{Anjan A. Reijnders}
\email[]{reijnder@physics.utoronto.ca}
\author{Jason Hamilton}
\author{Vivian Britto}
\affiliation{Department of Physics \& Institute for Optical Sciences, University of Toronto, 60 St. George Street, Toronto, ON M5S 1A7, Canada.}

\author{Jean-Blaise Brubach}
\author{Pascale Roy}
\affiliation{Synchrotron SOLEIL, L'Orme des Merisiers Saint-Aubin, 91192 Gif-sur-Yvette, France.}

\author{Quinn D. Gibson}
\author{R.J. Cava}
\affiliation{Department of Chemistry, Princeton University, Princeton, New Jersey 08544, USA}

\author{K. S. Burch}
\email[]{burchke@bc.edu}
\affiliation{Department of Physics, Boston College, 140 Commonwealth Ave, Chestnut Hill, Massachusetts 02467, USA}

\date{\today}

\begin{abstract}
Pb$_{0.77}$Sn$_{0.23}$Se is a novel alloy of two promising thermoelectric materials PbSe and SnSe that exhibits a temperature dependent band inversion below 300 K. Recent work has shown that this band inversion also coincides with a trivial to nontrivial topological phase transition. To understand how the properties critical to thermoelectric efficiency are affected by the band inversion, we measured the broadband optical response of Pb$_{0.77}$Sn$_{0.23}$Se as a function of temperature. We find clear optical evidence of the band inversion at $160\pm15$ K, and use the extended Drude model to accurately determine a $T^{3/2}$ dependence of the bulk carrier lifetime, associated with electron-acoustic phonon scattering. Due to the high bulk carrier doping level, no discriminating signatures of the topological surface states are found, although their presence cannot be excluded from our data.
\end{abstract}

\maketitle

\section{Introduction}
Thermoelectrics are a subject of great interest to industry for their many applications as macro and micro-scale Peltier coolers, their waste-heat energy recycling potential (heat-to-electricity conversion) in industrial and consumer products, and their application as highly sensitive compact and robust sensors.\cite{Zebarjadi:2012ul,Riffat:2003tf,2008NatMa...7..105S,DiSalvo:1999vv,2006AcPPA.108..609H} An academic interest has also persisted as the optimization of thermoelectric efficiency offers an intriguing challenge in which ostensibly mutually exclusive physical properties must be combined in one material for the greatest performance. This is illustrated by the figure of merit $ZT={\frac{TS^{2}\sigma}{\kappa}}$, where $S$, $\sigma$, and $\kappa$ are the Seebeck coefficient, the DC conductivity, and the thermal conductivity, respectively. For high thermoelectric efficiency, $S$ and $\sigma$ must therefore be simultaneously enhanced, while $\kappa$ is deliberately suppressed. This is a nontrivial task, considering the intimate link between the parameters that define $ZT$. For example, both $\sigma$ and $\kappa$ depend on the carrier lifetime $\tau$, while both $S$ and $\sigma$ are a function of the band dispersion. Hence, the improvement of each individual quantity could result in an overall decrease of $ZT$, thus making experimental techniques that can disentangle the common limitations of each quantity highly desirable. 

Currently, the primary methods of studying thermoelectrics are by DC transport, optics, thermopower experiments and neutron scattering,\cite{Liang:2013tk,Delaire:2011gr,1971PhRvB...3.4299D,1967PhRv..157..608S,1990JPCM....2.2935B,Gibbs:2013hq,Ekuma:2012gl,Zhao:2014uh,1999BrJPh..29..771F,2013ApPhL.103z2109G} where the standard Drude model is often employed to extract temperature dependent information about carrier lifetimes. A major shortcoming of the Drude model, however, is the assumption that the carrier lifetime is frequency independent. When analyzing optical data, this often results in poor quality Drude fits, and carrier scattering rates with large uncertainties. 
A solution to this shortcoming is the application of the {\it extended} Drude analysis to a sample's optical response.\cite{2009NatPh...5..647Q} While commonplace in the optical investigation of strongly correlated systems,\cite{2005RvMP...77..721B,2009NatPh...5..647Q,Basov:2011ht,2012PNAS..10919161N,2007JPCM...19l5208H} the extended Drude model offers some benefits previously unexplored in the context of thermoelectrics. Using this approach grants access not only to the temperature dependence of the carrier lifetime, but also to its frequency dependence, shedding light on the various scattering mechanisms that govern $\sigma$ and $\kappa$. Moreover, with both high temperature and energy resolution,\cite{Mirzaei:2013cx} broadband (far infrared through visible) optical spectroscopy provides direct access to the free carrier plasma frequency and band mass, both affecting $S$ and $\sigma$, as well as spectral weight transfers and band transitions, useful for studying band gaps and band inversions.

To illustrate the versatility of this approach we've performed our measurements on Pb$_{0.77}$Sn$_{0.23}$Se, a compound combining the thermoelectric characteristics of its thoroughly studied binary constituents PbSe and SnSe\cite{Shukla:2007tt,Ekuma:2012gl,Gibbs:2013hq,1990JPCM....2.2935B,1967PhRv..157..608S,Zhao:2014du,Anand:2014vn} However, unlike its constituents, previous works on Pb$_{0.77}$Sn$_{0.23}$Se crystals have shown evidence of a temperature dependent band inversion, although neither the exact transition temperature, nor the temperature dependence of the gap has been identified conclusively. This is largely a result of limited energy and temperature resolution of the experimental probes that were used, with various reports ranging between 80 K and 200 K. \cite{1971PhRvB...3.4299D,2012NatMa..11.1023D,Wojek:2013hx,1967PhRv..157..608S,Liang:2013tk,Xi:2014uz} Hence, broadband optical spectroscopy provides a unique opportunity to study, in a very controlled manner, how carrier dynamics change throughout the inversion. Moreover, many traditional thermoelectric probes, such as transport and Seebeck measurements, rely on band structure models to extract the scattering rate. Such band structure assumptions are severely complicated by the band inversion, further motivating the investigation of Pb$_{0.77}$Sn$_{0.23}$Se by optical spectroscopy as features of the band inversion and the carrier lifetime can be measured independently in a single experiment. This discriminating quality becomes even more evident by considering previous transport studies of the Pb$_{1-x}$Sn$_{x}$Se system. While Pb$_{1-x}$Sn$_{x}$Se alloys have been studied for 50 years,\cite{Butler:1966vb} only specially prepared low-carrier extrinsic p-type Pb$_{0.77}$Sn$_{0.23}$Se was found to show abrupt changes in the DC transport properties across the band inversion.\cite{1971PhRvB...3.4299D} Interestingly, as-grown n-type crystals showed no clear signatures of the band inversion at all. A recent paper by Liang et al. highlighted this peculiarity and was the first to show evidence of the band inversion in n-type Pb$_{0.77}$Sn$_{0.23}$Se, employing Nernst and thermopower experiments.\cite{Liang:2013tk} Particularly for device applications, a thorough understanding of the electronic properties of both n and p-type crystals is pivotal.\cite{Anand:2014vn} \\

The temperature dependent band inversion of Pb$_{0.77}$Sn$_{0.23}$Se has also received significant attention from the topological insulator community. It was recently shown that while PbSe is a trivial insulator, SnSe is a topological crystalline insulator (TCI); a state of matter in which crystal point group symmetry protects the persistence of gapless surface states with a spin polarized Dirac dispersion.\cite{2011PhRvL.106j6802F,Xu:2012bm,2012NatMa..11.1023D,Xi:2014uz} Interestingly, in Pb$_{0.77}$Sn$_{0.23}$Se we have the unique opportunity to tune between topological and trivial phases by temperature, rather than doping, thus negating the effects of disorder. Moreover, since it is challenging to perform ARPES and STM (the primary methods of studying surface states) at various temperatures, temperature dependent broadband spectroscopy makes a great tool for the investigation of surface states in topological (crystalline) insulators.\cite{BTSpaper,2012PhRvL.108h7403V,Jenkins:2013vj,DiPietro:2013bp,Chapler,Wu:2013tj,2013PhRvB..88d5414L,2013PhRvL.111o5701X,Tran:2014if,Xi:2014uz,Anand:2014vn} We note that since band gaps are typically tuned by electron-phonon coupling and anharmonicity, the concurrent band inversion and topological phase transition hints at a more subtle process. Hence, Pb$_{0.77}$Sn$_{0.23}$Se can reveal new insights into how anharmonicity, electron-phonon coupling and topology all interplay.
 \\

In the present work, we measure the temperature dependent broadband reflectance of Pb$_{0.77}$Sn$_{0.23}$Se between 5 K - 292 K, and extract its optical conductivity and dielectric function between 6 meV - 6 eV. Subsequent extended Drude analysis reveals the frequency dependent free carrier scattering rate, and is used for a precise determination of the dominant temperature dependent scattering mechanism. We find clear optical evidence of the band inversion temperature at $160\pm15$ K, by studying the free carrier plasma frequency and direct optical band gap transitions. Interestingly, the temperature dependent free carrier scattering rate shows no signatures of the band inversion, and follows $T^{3/2}$ behaviour in accordance with electron-acoustic phonon scattering. Finally, we find that the Fermi level lies deep in the conduction band ($114\pm22$ meV), obscuring the observation of surface states,\cite{BTSpaper,2012PhRvB..86d5439D,LaForge:2010dx,Akrap:2012fy} and causing the Seebeck coefficient and power factor to be insensitive to the band inversion.


\section{Experiment}
%
\begin{figure}
\includegraphics[scale=0.31]{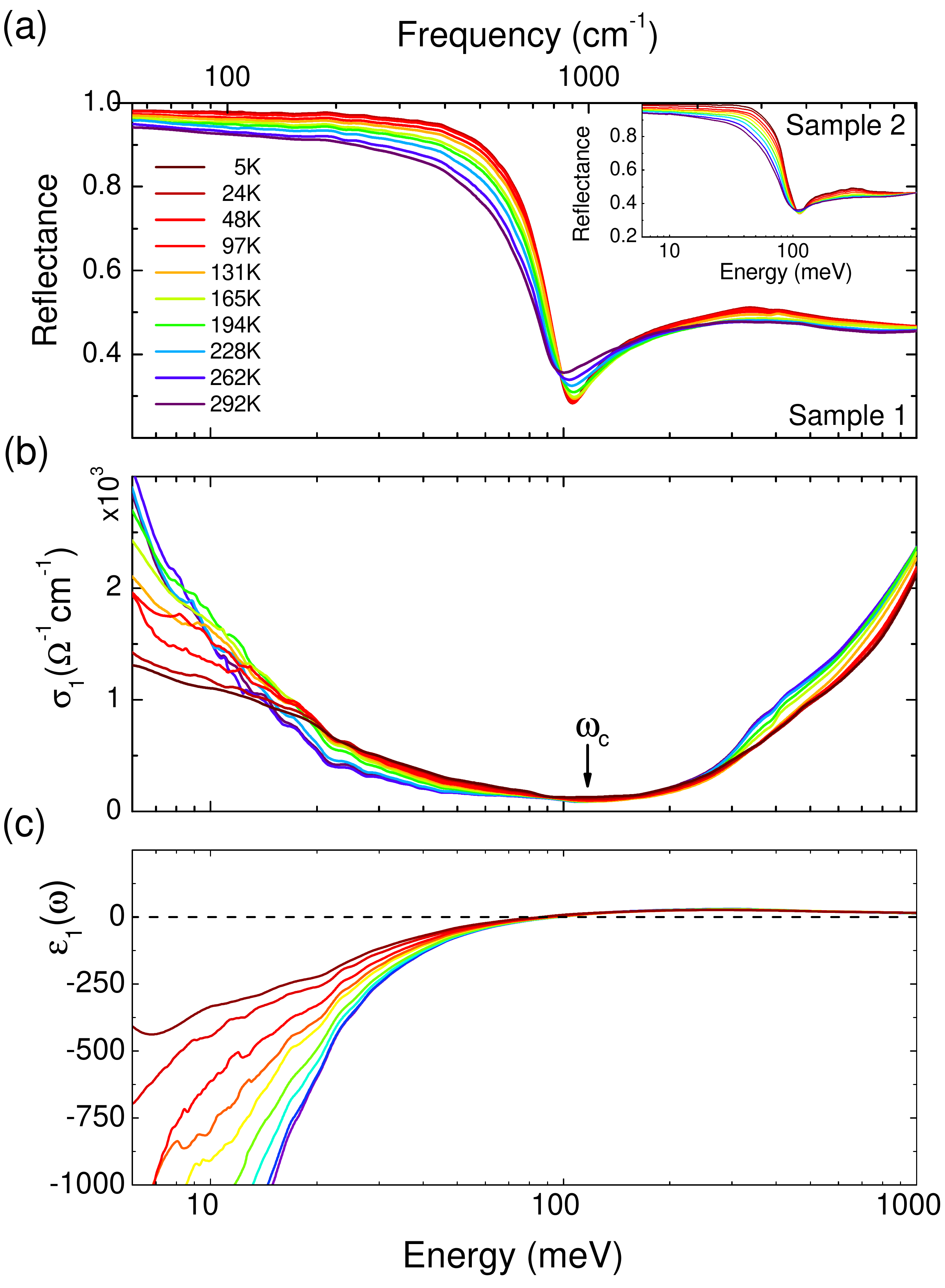}
\caption[Reflectance, real optical conductivity, and real dielectric function of Pb$_{0.77}$Sn$_{0.23}$Se]{\label{fig:PSSFig1}(a) The changing slope in the reflectance below the plasma edge (reflectance minimum) indicates a temperature dependent free carrier scattering rate in Pb$_{0.77}$Sn$_{0.23}$Se. (b) The optical conductivity reveals a clear separation ($\omega_c$) between Drude conductivity below 125 meV, and inter/intra-band transitions above 125 meV. (c) The zero-crossing of the real part of the dielectric function ($\epsilon_1$), indicative of the screened plasma frequency, can be used to determine $\epsilon_{\infty}$}
\end{figure}
%
Pb$_{0.77}$Sn$_{0.23}$Se samples were grown at Princeton University.\cite{Liang:2013tk} Prior to the reflectance measurements, samples were freshly cleaved with a razor blade to produce a lustrous surface. Cleaving by tape was also attempted, but proved unsuccessful in producing high quality surfaces.
Temperature dependent reflectance measurements were performed in near-normal geometry between 6 meV - 1.5 eV, using a modified Bruker VERTEX 80v and ARS continuous flow cryostat, described in detail elsewhere\cite{BTSpaper} (Fig.~\ref{fig:PSSFig1}a). Reflectance data below 12 meV were corroborated by repeating the measurements using synchrotron radiation and a custom designed sample chamber with a closed cycle Cryomech cryostat at the AILES Beamline of the SOLEIL synchrotron in France. All infrared data were complemented by partially overlapping room temperature ellipsometry measurements between 0.75 meV - 6 eV. 

Using RefFit, a simultaneous variational dielectric fit (VDF) of Reflectance and ellipsometry data yielded all Kramers-Kronig consistent optical constants. All measurements were performed on two different crystals from the same sample batch, showing excellent qualitative resemblance (see inset Fig.~\ref{fig:PSSFig1}a for the reflectance of sample 2). Hence, data from only one crystal are discussed.

\section{Results and Analysis}
Fig.~\ref{fig:PSSFig1}b shows the real part of the optical conductivity $\sigma_1(\omega)$ of Pb$_{0.77}$Sn$_{0.23}$Se in the far and mid infrared. The spectra are dominated by a distinct Drude peak below 125 meV, associated with free carrier conductivity, and separated by a cut-off frequency $\omega_{c}$ from the interband transitions above 125 meV. 

The simplest approach to understanding the spectral range below $\omega_{c}$ is to apply the semi-classical Drude model, in which the complex dielectric function for the free carrier response is given by 
\begin{equation}\label{eqn:Drude}
\hat\epsilon(\omega) = \epsilon_{\infty}-\frac{\omega_p^2}{\omega^2+i\frac{\omega}{\tau}}
\end{equation}
where $\omega_p$ is the bare plasma frequency, $\tau$ is the transport lifetime, and $\epsilon_{\infty}$ captures all contributions to the dielectric function other than the Drude conductivity (i.e. interband transitions and polarizability of the static background). The bare plasma frequency is defined as $\omega_p^2=4\pi ne^2/m_{b}$, where $n$ is the carrier density and $m_{b}$ is the band mass. 

An immediate consequence of this model is the large, low frequency negative contribution to the real part of the dielectric function ($\hat\epsilon=\epsilon_1+i\epsilon_2$) due to the free carrier response. Indeed from eq.~\ref{eqn:Drude} it can be seen that in the limit $\omega\tau\gg1$ the zero-crossing of $\epsilon_1$ corresponds to the screened plasma frequency $\tilde{\omega}_{p}$, which is related to the bare plasma frequency $\omega_{p}$ through $\omega_{p} = \tilde{\omega}_{p}\sqrt{\epsilon_\infty}$. Hence, we expect to observe this zero-crossing in $\epsilon_1$, which can be seen directly in \ref{fig:PSSFig1}c. The temperature dependence of both $\omega_p$ and $\epsilon_{\infty}$ are discussed next.

\begin{figure}
\includegraphics[scale=0.40]{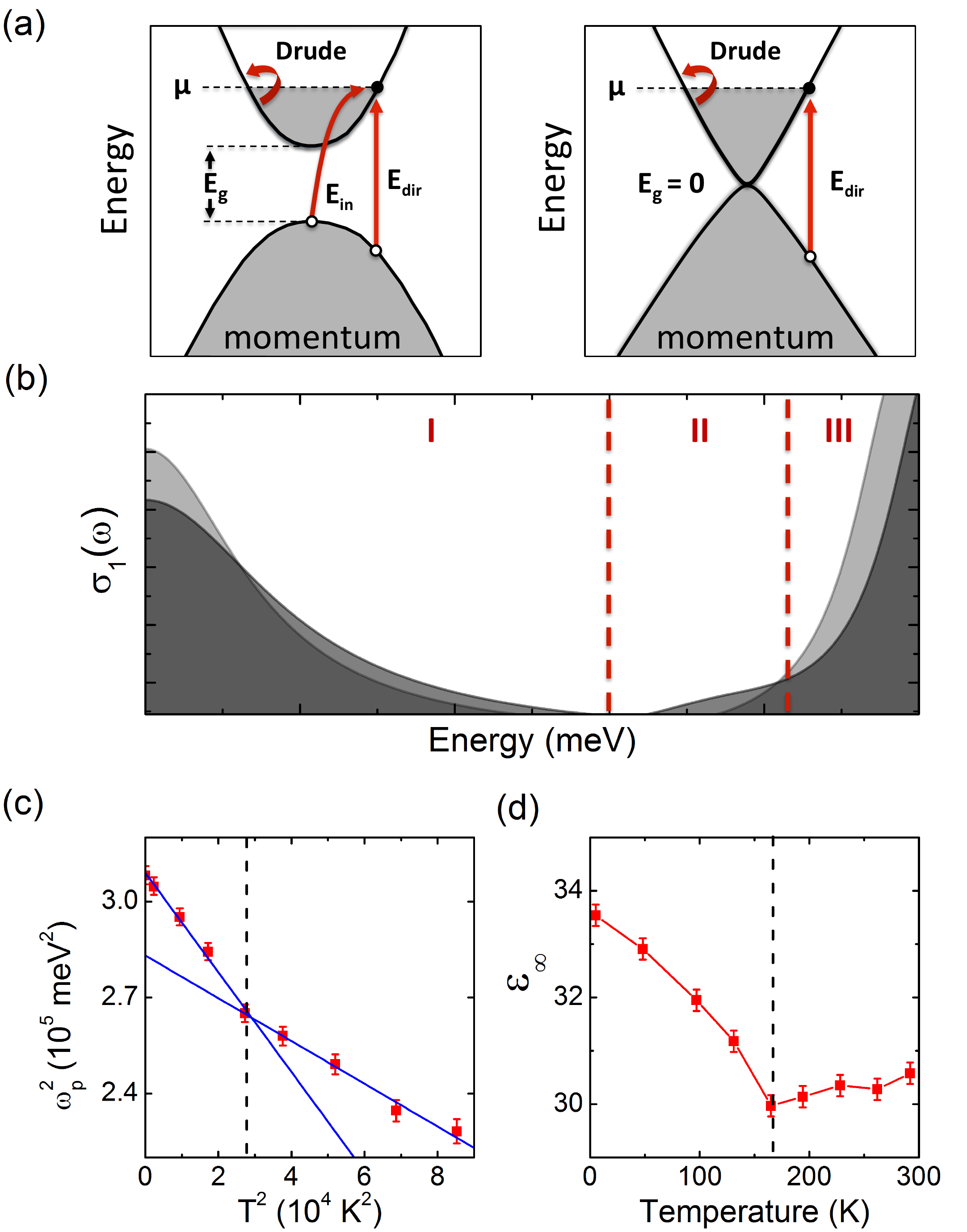}
\caption[Band structure cartoon, free carrier spectral weight, and $\epsilon_{\infty}$ of Pb$_{0.77}$Sn$_{0.23}$Se]{\label{fig:PSSFig2} (a) Band structure cartoon before or after the band inversion (left panel), and during the inversion (right panel), illustrating the allowed (in)direct transitions when the chemical potential is in the bulk conduction band (as is the case in as-grown Pb$_{0.77}$Sn$_{0.23}$Se). (b) Signatures of this inversion include a changing free carrier lifetime $\tau(T,\omega)$ and plasma frequency $\omega_p$ (region I), spectral weight change for indirect gap transitions (region II), and changes in the size of the direct gap (region III). (c) The plasma frequency of Pb$_{0.77}$Sn$_{0.23}$Se shows a distinct change in temperature dependence around 160 K, deviating from it's expected quadratic temperature dependence.  (d) $\epsilon_\infty$ also shots an abrupt change in its temperature dependence at 160 K, further corroborating 160 K as the band inversion temperature.}
\end{figure}

Before analyzing the spectra in detail, it is useful to consider what we expect to observe as a result of the band inversion. For compounds in which the chemical potential is located in the bulk gap, a distinct signature of a band inversion would be the spectral weight redistribution from interband transitions to the Drude conductivity. As the gap closes, the onset frequency of the interband transitions (roughly $\omega_c$ in Fig.~\ref{fig:PSSFig1}c) moves to lower energy, until it merges with the Drude when the gap size reduces to zero. However, since stoichiometric Pb$_{0.77}$Sn$_{0.23}$Se is naturally n-type (chemical potential in the conduction band, shown in Fig.~\ref{fig:PSSFig2}a), we do not expect to observe such obvious behaviour, and therefore perform a more sophisticated analysis to observe the band inversion. 
%
%
%
Fig.~\ref{fig:PSSFig2}b illustrates how the band inversion could be deduced from signatures in $\sigma_1(\omega)$.  Region I is dominated by free carrier behaviour, described by eqn.~\ref{eqn:Drude}. Hence, as the band dispersion transforms across the band inversion (Fig.~\ref{fig:PSSFig2}a), changes in both $m_b$ and $n$ can result in an unusual temperature dependence of the plasma frequency (where $\omega_p^2\propto$ spectral weight in $\sigma_1(\omega)$). One might also expect the inversion driven changes in the electronic structure to be reflected in the carrier lifetime. This can be seen from the width of the Lorentzian features (centred at 0), while deviations from the Lorentzian lineshape indicate an energy dependent lifetime. Region III can be used to extract the magnitude of the direct gap transitions $E_{dir}$  shown in Fig.~\ref{fig:PSSFig2}a. With the chemical potential ($\mu$) in the conduction band, the smallest allowed $E_{dir}$ is offset from the conduction band minimum. Hence, changes in the band dispersion as well as $n$ are likely to affect $E_{dir}$, resulting in a shift of the interband transition onset in $\sigma_1(\omega)$, as shown in Fig.~\ref{fig:PSSFig2}b. Since $\epsilon_{\infty}$ is a measure of the spectral weight in $\sigma_1$ above the free carriers, this quantity too should reveal signatures of the band inversion, similar to $E_{dir}$. Finally, below $E_{dir}$, region II is sensitive to temperature dependent changes in indirect gap transitions $E_{in}$, where $E_{in}<E_{dir}$, as well as impurity scattering responsible for an extended $E_{dir}$ onset (known as an Urbach tail\cite{1989PhRvB..39.1140G}).

\subsection{Bulk plasma frequency and $\epsilon_{\infty}$}
\label{sec:PSSbulk}

The spectral weight of the Drude feature (area under $\sigma_1$) is a direct measure of the bare plasma frequency squared, $\omega_p^2$, and was determined by integrating $\sigma_1(\omega)$ up to a cut-off point $\omega_{c}$ using
%
%
\begin{equation}\label{eqn:Wp}
\omega_{p}^2 = \frac{120\Omega}{\pi}\int_{0}^{\omega_{c}} \! \sigma_{1}({\omega'}) \, \mathrm{d}\omega'
\end{equation}
%
%
where $\omega_{c}$ separates the Drude contribution to $\sigma_{1}(\omega)$ from the interband transitions (as indicated in Fig.~\ref{fig:PSSFig1}b).\cite{2007JPCM...19l5208H} The bare plasma frequency was then used to extract $\epsilon_\infty$ from the zero-crossing in $\epsilon_1(\omega)$. We note that an alternative method to find $\epsilon_\infty$ (which does not rely on the zero-crossing and is described elsewhere\cite{2007JPCM...19l5208H}) was also used, and resulted in similar temperature dependent values differing only by a constant offset of a few percent. 

The temperature dependence of $\omega_p^2$ and $\epsilon_\infty$ are shown in Fig.~\ref{fig:PSSFig2}c and \ref{fig:PSSFig2}d, respectively. Starting with $\omega_p^2$, we find a reduction as temperature increases, with a distinct discontinuity in the trend around 160 K. In general, a temperature dependent reduction in $\omega_p^2$ is an expected result that can be understood as follows. While semi-classically we have $\omega_p^2\propto n/m$, quantum mechanically the squared plasma frequency calculated using eqn.~\ref{eqn:Wp} corresponds to an integration of the density of states ($N(\omega,T)$) multiplied by the Fermi function ($f(\omega,T)$), so that $\omega_p^2(T)\propto\int \! N(\omega,T)f(\omega,T)\, \mathrm{d}\omega$.\cite{Benfatto:2005km,Carbone:2006iw} Evaluating this integral using the Sommerfeld expansion yields a temperature independent term equal to $\omega_p^2$ at 0 K, and a negative $T^2$ term as a result of thermal smearing of the Fermi function. Hence, the temperature dependence of the squared plasma frequency is expected to follow
%
%
\begin{equation}\label{eqn:sommerfeld}
\omega_{p}^2(T) \approx \omega_{p}^2(T=0K) - BT^2
\end{equation}
%
%
where $B$ depends on $N(\omega,T)$ at the Fermi energy.\cite{Benfatto:2005km,2005PhRvL..94f7002O,Carbone:2006iw} Interestingly, while a linear decline in $\omega_p^2$ vs $T^2$ is expected, Fig.~\ref{fig:PSSFig2}c shows two distinct regimes in its temperature dependence below and above $160\pm15$ K. Since $B$ is a function of the density of states, this is a clear signature of an abrupt change in the band dispersion, associated with the band inversion. Indeed, this critical temperature falls within the 80 K - 200 K temperature range over which the band inversion was previously observed.\cite{1971PhRvB...3.4299D,2012NatMa..11.1023D,Wojek:2013hx,2013arXiv1306.0043G,1967PhRv..157..608S,Liang:2013tk} This also provides a first indication that despite the Fermi level of n-type Pb$_{0.77}$Sn$_{0.23}$Se not being in the gap, discontinuities in its optical properties are a useful probe in the exact determination of its band inversion temperature.

Besides clear signatures of the band inversion in $\omega_p^2$, we also expect to see a change in $\epsilon_{\infty}$ at $160\pm15$ K. Since $\epsilon_{\infty}$ is a measure of the spectral weight of all interband transitions, it is sensitive to temperature dependent changes in the band structure.\cite{2007NJPh....9..229K} This is confirmed by (Fig.~\ref{fig:PSSFig2}d), revealing a clear change in the temperature dependence around 160 K, and will be discussed in greater in section~\ref{sec:PSSinterband}.

\subsection{Surface States}
\label{sec:PSSss}
Besides the Seebeck sensitivity to the band inversion at $160\pm15$ K, this critical temperature is also relevant to the topological phase of Pb$_{0.77}$Sn$_{0.23}$Se. Previous studies have shown how the band inversion is accompanied by a topological phase transition, where at low temperature topologically protected surface states persist in the bulk band gap.\cite{2011PhRvL.106j6802F,2012NatMa..11.1023D,Wojek:2013hx,2013arXiv1306.0043G,2013arXiv1305.2823O,Liang:2013tk} In our optical data below 160 K (Fig.\ref{fig:PSSFig1}a) we see no evidence of Dirac surface states, such as previously observed in topological insulators Bi$_2$Te$_2$Se, Bi$_2$Se$_3$, and (Bi,Sb)$_2$Te$_3$\cite{BTSpaper,2012PhRvL.108h7403V,Jenkins:2013vj,DiPietro:2013bp,Chapler,Wu:2013tj} This is not surprising since we find a bulk chemical potential of $114\pm22$ meV above the conduction band minimum at 5 K (shown later), consistent with previous studies of Pb$_{1-x}$Sn$_{x}$Se with similar stoichiometry\cite{2012NatMa..11.1023D,2013arXiv1306.0043G,Liang:2013tk,Wojek:2013hx} Such a high bulk $\mu$ causes bulk conductance to dominate the optical properties, which can by understood by considering how the measured reflectance relates to the complex dielectric function, as $R=|(1-\sqrt{\hat\epsilon})/(1+\sqrt{\hat\epsilon})|^2$. From Fig.~\ref{fig:PSSFig1}c and eqn.~\ref{eqn:Drude} it is clear that $|1\pm\sqrt{\hat\epsilon}|\gg1$ for all frequencies $\omega<\tilde{\omega}_p$, resulting in $R>95\%$ over the full far infrared (FIR) range for most temperatures. Such high FIR reflectance even above the topological phase transitions obscures discriminating surface conductance signatures from the bulk conductivity in $R(\omega)$. Moreover, with the Fermi level so high in the conduction band, mixing of the surface and bulk bands also blurs other distinguishing features such as the bulk vs surface carrier lifetime. However, to ensure we did not overlook surface state signatures in our reflectance data, we used the Boltzmann transport equation to estimate their expected effect on the measured reflectance. Indeed, the surface state signatures are far below our detection sensitivity, as described in the Appendix, and crystals with a lower bulk carrier concentration are required for further optical investigations of the topological crystalline surface state properties of Pb$_{0.77}$Sn$_{0.23}$Se.

\subsection{Interband transitions}
\label{sec:PSSinterband}
\begin{figure}
\includegraphics[scale=0.30]{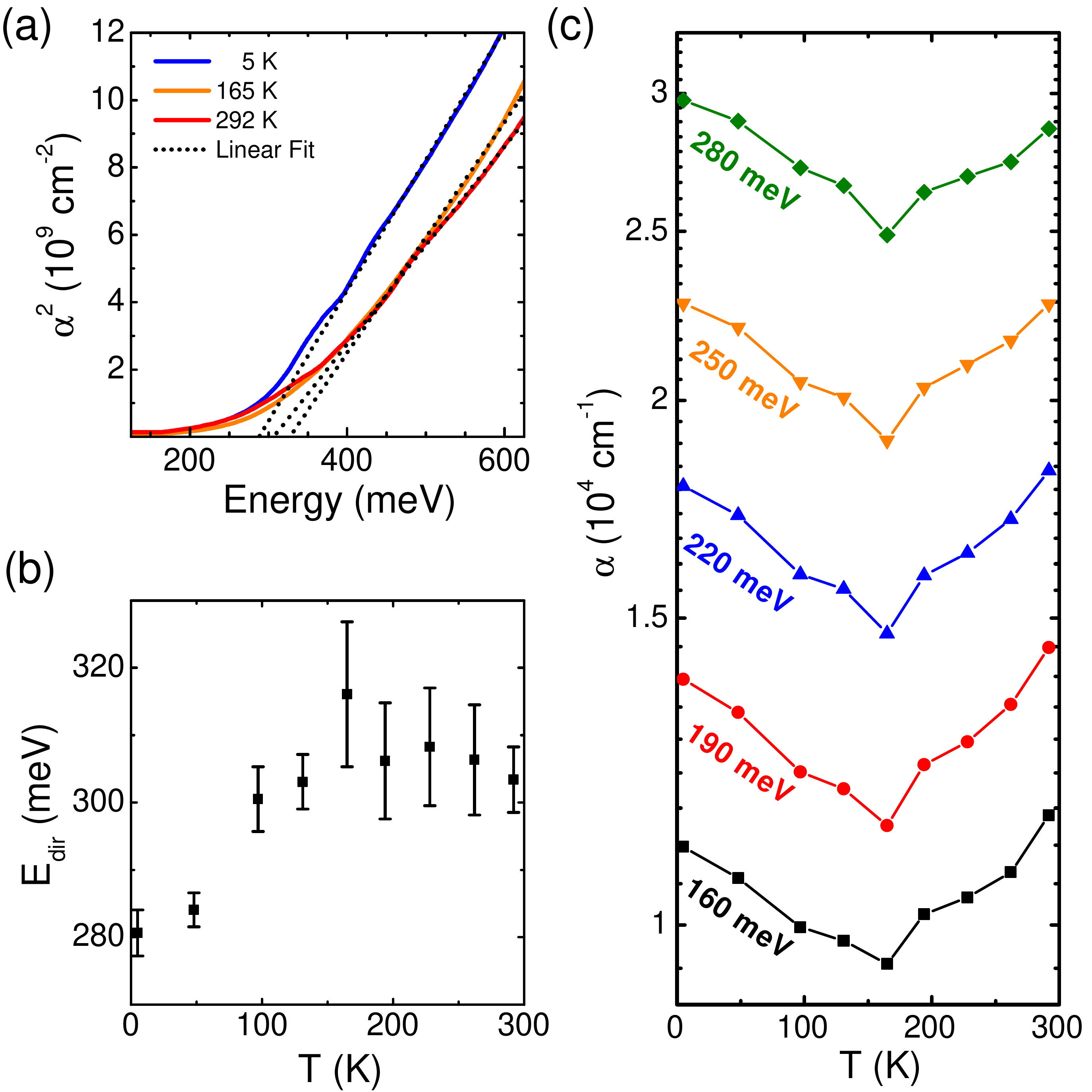}
\caption[Linear fit of $\alpha^2$, the direct gap E$_{dir}$(T), and $\alpha$(T) at various energies below the direct gap of Pb$_{0.77}$Sn$_{0.23}$Se]{\label{fig:PSSFig3} (a) Linear fits of the squared absorption coefficient yield the smallest observable direct band transition energy (x-intercept). (b) Direct gap transition extracted from the linear fit. (c) Absorption coefficient (log scale) below $E_{dir}$, revealing a distinct change in temperature dependence, uncharacteristic for an Urbach tail.}
\end{figure}

To study how the bulk electronic structure is affected by the band inversion, we turn to the band gap transitions that are visible above $\omega_{c}$ in Fig.~\ref{fig:PSSFig1}b. With ARPES measurements showing $\mu$ in the conduction band for crystals with identical stoichiometry,\cite{2012NatMa..11.1023D,Wojek:2013hx,2013arXiv1306.0043G} the smallest allowed direct gap transition exceeds the direct bandgap (and thus takes place away from the Brillouin Zone centre), as illustrated in Fig.~\ref{fig:PSSFig2}a. This effect is known as the Burstein-Moss shift.\cite{YuCard,Gibbs:2013hq} Assuming a parabolic dispersion for both the conduction and valence band, the energy of this transition can be estimated. By adding $\mu=\hbar^2 k_F^2/2m_c$ to the direct band gap energy $E_g$, and to the energy of the valence band electron with respect to valence band maximum of $E_v=\hbar^2 k_F^2/2m_v$, the total transition energy at zero energy is given by 
%
%
\begin{equation}\label{eqn:directgap}
E_{dir}\approx E_g+\mu\left(1+\frac{m_c}{m_v}\right)
\end{equation}
%
%
The magnitude of $E_{dir}$ can be obtained from a linear fit of the squared absorption coefficient, where $\alpha^2=C(\omega-E_{dir})$, as shown in Fig.~\ref{fig:PSSFig3}a.\cite{YuCard} The temperature dependence of $E_{dir}$ is shown in Fig.~\ref{fig:PSSFig3}b, where a clear maximum of $315\pm10$ meV is observed around 160 K. 
A previous study of the doping dependent band inversion in the Pb$_{1-x}$Sn$_x$Te system also found a change in sign of $dE_{dir}/dT$ as the bands invert.\cite{1999BrJPh..29..771F} Interestingly, for composition Pb$_{0.44}$Sn$_{0.56}$Te, in which a temperature dependent inversion takes place, a minimum in $E_{dir}$ is observed at the inversion temperature, in contrast to our maximum.  Since the direct gap $E_g$ is zero at $160\pm15$ K, eqn.~\ref{eqn:directgap} suggests that this maximum is either due to an increase in $\mu$, or a change in ${m_c}/{m_v}$ across the transition. Both scenarios are plausible, and require further study specifically addressing the temperature dependent band mass and $\mu$. 

We find a minimum direct transition of 280 meV at 5 K, which is larger than the low temperature direct transition of 200-250 meV shown by previous ARPES results.\cite{2012NatMa..11.1023D,Wojek:2013hx,2013arXiv1306.0043G} This difference is expected considering our larger carrier density compared to previous reports, as we show later in section {\bf DC Transport}. We also note that the sensitivity to surface states in ARPES conceals bulk signatures, complicating the estimation of bulk transition energies.
%
%

We now consider the spectral weight below the direct transition energy $E_{dir}$. This is often explained as an Urbach tail, which results from disorder in the crystal, and has an exponential energy dependence that changes monotonically with temperature.\cite{1989PhRvB..39.1140G,Gibbs:2013hq} While Fig.~\ref{fig:PSSFig3}a does show exponential behaviour below $E_{dir}$ ($\sim280$ meV), a closer look at $\alpha(T)$ at various frequencies in Fig.~\ref{fig:PSSFig3}c reveals atypical Urbach behavior. Instead of a monotonic temperature dependent change of the exponential behaviour, a distinct minimum is observed at the 160 K band inversion temperature, while at both lower and higher temperatures the spectral weight in the absorption coefficient increases. Although it is not surprising that transitions below the direct gap are sensitive to the band inversion, the exact details of their temperature dependence are not clear and require future theory/experiment. Nonetheless, this shows that the absorption/optical conductivity alone could be used to detect the transition. 

\subsection{Extended Drude analysis and scattering rate}

\begin{figure}
\includegraphics[scale=0.31]{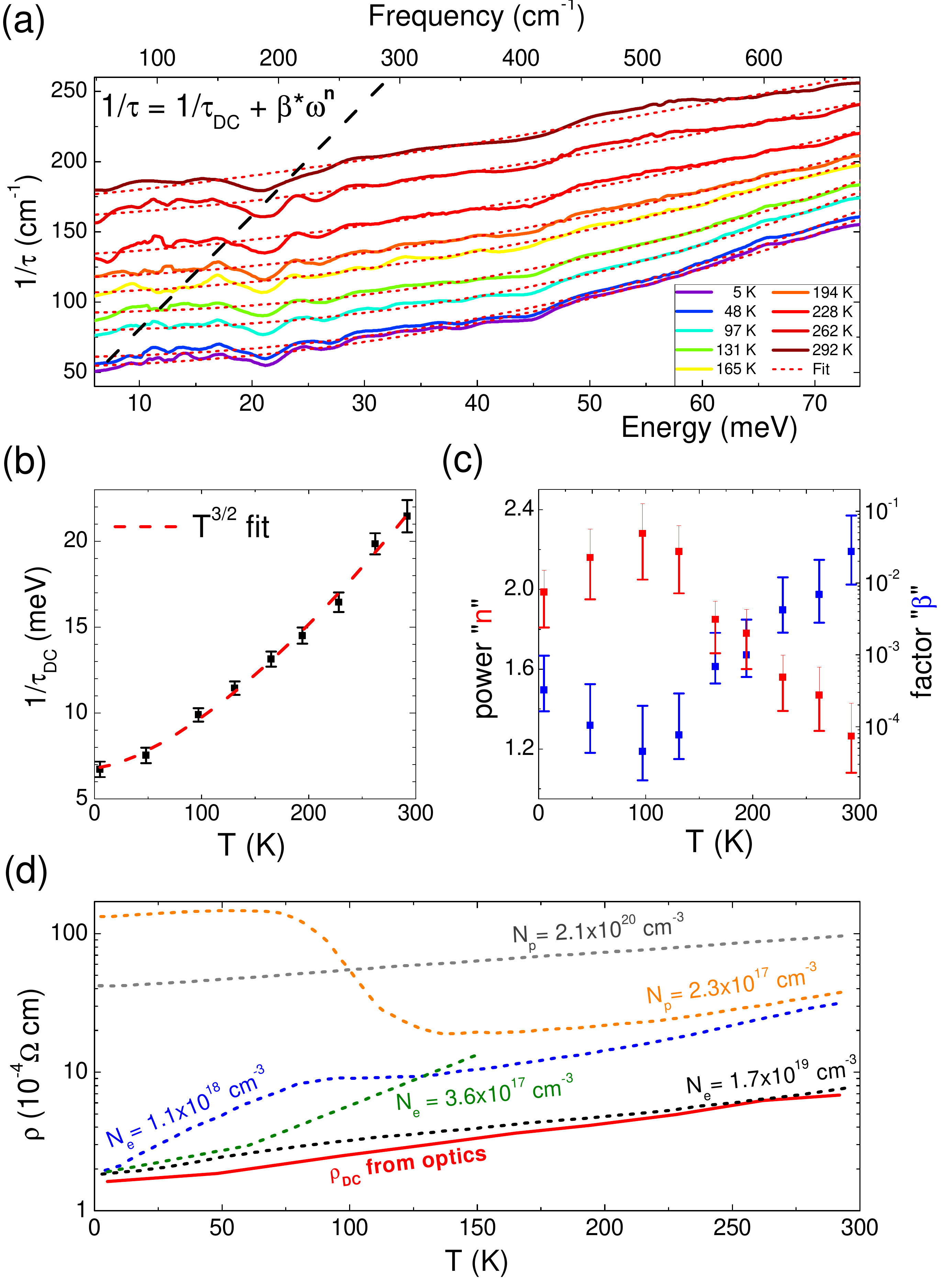}
\caption[The frequency dependent scattering rate, DC scattering rate, fitting parameters and the resistivity of Pb$_{0.77}$Sn$_{0.23}$Se]{\label{fig:PSSFig4} (a) Allometric fits of the frequency dependent scattering rate. The black dashed line corresponds to $\omega=1/\tau$. (b) A T$^{3/2}$ fit (red dashed line) of the temperate dependent static scattering rate $1/\tau_{DC}$ illustrates phonon dominated scattering. (c) While it's difficult to interpret fitting parameters $n$ and $\beta$ due to the small frequency range over which $1/\tau$ could be fitted, they both reveal an abrupt change in temperature dependence around the band inversion temperature $\sim$160 K. (d) Comparison of Pb$_{0.77}$Sn$_{0.23}$Se resistivity obtained by optical spectroscopy (solid red line), and by standard DC transport methods by Liang et al.\cite{Liang:2013tk} (green dotted line), and Dixon et al.\cite{1971PhRvB...3.4299D} (all other data).}
\end{figure}

To study the bulk carrier lifetime in Pb$_{0.77}$Sn$_{0.23}$Se, we now turn to the extended Drude model. From eqn.~\ref{eqn:Drude}, it can be seen that the free carrier response in the standard Drude model is fully defined by $\epsilon_{\infty}$, $\omega_p$, and $\tau$, where $\tau$ is the static transport lifetime resulting from a sum of all scattering processes according to Matthiessen's rule (including impurity, electron-phonon and electron-electron scattering).\cite{singleton} While the Drude model is frequently used to study scattering mechanisms that can be distinguished by their temperature dependence, the temperature dependent band inversion in Pb$_{0.77}$Sn$_{0.23}$Se severely complicates such analysis. Moreover, the standard Drude model does not consider the energy dependence of the lifetime $\tau(\omega)$. 
Since the carrier lifetime is an important parameter for thermoelectric efficiency\cite{Jonson:1980tc} we turn to the extended Drude model, in which the lifetime has an energy dependence given by\cite{2009NatPh...5..647Q} 
%
%
\begin{equation}\label{eqn:1overtau}
\frac{1}{\tau(\omega)} = -\frac{\omega_p^2}{\omega}\mathrm{Im}\left(\frac{1}{\hat{\epsilon}(\omega)-\epsilon_{\infty}}\right)
\end{equation}
%
%
Fig.~\ref{fig:PSSFig4} shows the frequency dependent scattering rate obtained using eqn.~\ref{eqn:1overtau}. The black line shows where $\omega=1/\tau$, below which the quasiparticle transport is considered to be coherent.\cite{2009NatPh...5..647Q} To study the nature of the dominant scattering mechanism, the data were fitted (red dotted lines in Fig.~\ref{fig:PSSFig4}a) using the empirical relation
\begin{equation}\label{eqn:allometric}
\frac{1}{\tau(\omega)}=\frac{1}{\tau_{DC}}+\beta\omega^n
\end{equation}
where $1/\tau_{DC}=\alpha T^m+1/\tau_{imp}$.  Here, $1/\tau_{imp}$ accounts for impurity scattering, and $\alpha$ and $\beta$ are constants that depends on material properties.
When $n=m=2$, and for certain values of $\alpha/\beta$, eqn.~\ref{eqn:allometric} resembles the single-particle scattering rate within Fermi Liquid theory, in which electron-electron scattering dominates the transport properties. \cite{2012PNAS..10919161N} 

Fig.~\ref{fig:PSSFig4}b shows $1/\tau_{DC}$, obtained from the fits with eqn.~\ref{eqn:allometric}. As the temperature is raised, $T^{3/2}$ behavior is observed (dashed red line), associated with electron-acoustic phonon scattering in semiconductors.\cite{YuCard,singleton} This is consistent with most good thermoelectrics, in which electron-acoustic phonon scattering dominates the transport properties.\cite{Pei:2011kx,Delaire:2011gr,Wang:2011ip,1971PSSBR..43...11R} Perhaps somewhat unexpected, however, is that the band inversion appears to have no bearing on $1/\tau_{DC}$. 
This is contrasted by recent work by L. Wu et al.,\cite{Wu:2013tj} in which the scattering rate of carriers in (Bi$_{1-x}$In$_{x}$)$_2$Se$_3$ was found to diverge going across the topological phase transition as a function of doping. 

We note that since we are most sensitive to bulk carriers (with $\mu$ high in the conduction band) we don't expect to see such a divergence, but it is worth emphasizing that more resistive Pb$_{1-x}$Sn$_{x}$Se samples (in which the surface contributions are enhanced) may exhibit similar behaviour. For such samples, the frequency dependent scattering rate may also serve as a novel analytical technique to discriminate surface states from the bulk. Since surface states are protected from backscattering, their lifetimes are expected to exceed the bulk carrier lifetime. This could result in distinct surface and bulk scattering contributions in $1/\tau(\omega)$, for which we see no evidence in our samples. From fitting $1/\tau_{DC}$, we also find a constant $1/\tau_{imp}=6.8\pm0.3$ meV, attributed to impurity scattering. We note that in metals and degenerate semiconductors such as as-grown Pb$_{0.77}$Sn$_{0.23}$Se, the electronic impurity scattering rate $1/\tau_{imp}^{\sigma}$ is roughly equal to the thermal impurity scattering lifetime $1/\tau_{imp}^{\kappa}$.\cite{singleton}

Fig.~\ref{fig:PSSFig4}c shows the temperature dependence of fitting parameters $n$ and $\beta$. Since we were only able to obtain reliable $1/\tau(\omega)$ data over a limited range spanning just one decade in energy (as seen in Fig.~\ref{fig:PSSFig4}a), we hesitate to attribute much physical meaning to either parameter. Additional measurements at lower frequencies and temperatures beyond the scope of this work, or detailed studies by alternative probes such as high resolution temperature dependent ARPES could offer greater insights. Nevertheless, a considerable discontinuity in both $n$ and $\beta$ at 160 K, which does not fall within our error bars, does suggest that the frequency dependent scattering rate is tied to the band inversion.  Moreover, with $n=2$ normally associated with Fermi liquid type electron-electron dominated scattering,\cite{2012PNAS..10919161N} this high value of $n$ is quite surprising considering the relatively high temperatures and frequencies up to which this behavior persists, making it an interesting starting point for followup studies.

\subsection{DC Transport}
\label{sec:DCtransport}

For comparison with previous transport studies of Pb$_{0.77}$Sn$_{0.23}$Se, we calculate its DC conductivity using $\sigma_{o}=\omega_{p}^{2}\tau_{DC}/60$ and find $\sigma_{o}=6.2\times10^{3}$ $\Omega^{-1}$cm$^{-1}$ at 5 K. This low temperature value is in good agreement with previous transport results of n-type samples with identical stoichiometry ($6.7\times10^{3}$ $\Omega^{-1}$cm$^{-1}$ by Dixon et al.\cite{1971PhRvB...3.4299D} and $5.5\times10^{3}$ $\Omega^{-1}$cm$^{-1}$ by Liang et al.\cite{Liang:2013tk}). A comparison of the resistivity with previously published data is shown in Fig.~\ref{fig:PSSFig4}d, where the dashed lines are digitized data from Liang et al.\cite{Liang:2013tk} (green) and Dixon et al.\cite{1971PhRvB...3.4299D} (all other colors). We note that all n-type crystals increase in resistance with temperature, typical for metals and degenerate semiconductors. Only the resistivity of a low carrier p-type sample measured by Dixon et al.\cite{1971PhRvB...3.4299D} shows an anomaly between 75 K and 200 K, possibly associated with the band inversion temperature, while none of the n-type curves reveal such signatures. This further illustrates the utility of optical spectroscopy, which can simultaneously determine $\omega_p$ and $1/\tau$ to find $\sigma_0$, while also measuring E$_F$ (via direct transitions) and the inversion temperature. 

A possible explanation for the difference between n-type and p-type resistivity behavior is the asymmetry of the conduction and valance band, which can be tested by comparing their respective band masses. Using the carrier density of $N_{e}=1.7\times10^{19}$ cm$^{-3}$ (from the resistivity curve most similar to our our results), we find a mobility of $2260$ cm$^{2}$ V$^{-1}$ s$^{-1}$, and a band mass of $m_{b}=0.076$m$_e$, using $\sigma_{o}=4\pi n e^2\tau(\omega)/m_{b}=ne\mu$. This band mass is smaller than the value of 0.1m$_e$ found by Dixon et al. for p-type Pb$_{0.77}$Sn$_{0.23}$Se, thus confirming the asymmetry between conduction and valence band. Using eqn.~\ref{eqn:directgap} we can now estimate the value of $\mu$ ($\approx E_F$), using the low temperature direct gap value of E$_g=80\pm20$ meV, consistent with previous studies and calculations.\cite{1967PhRv..157..608S,2012NatMa..11.1023D,Wojek:2013hx} We find a value for $\mu$ of $114\pm22$ meV. To further corroborate this value, we derive an expression for E$_F$ using the parabolic band approximation. Starting with E$_F=\hbar^2 k_F^2/2m_b$, and substituting the 3D carrier density $n_{3D}=\tfrac{1}{(2\pi)^3}\tfrac{4}{3}\pi k_F^3g_s g_v$, we get
%
%
\begin{equation}\label{eqn:Efder1}
E_F=\frac{\hbar^2}{2m_b}\left(\frac{6\pi^2 n}{g_s g_v}\right)^{2/3}
\end{equation}
%
%
where $g_s=2$ and $g_v=4$ are the bulk spin and valley degeneracy, respectively.\cite{2012NatMa..11.1023D,Liang:2013tk} Substituting the plasma frequency in SI units into eqn.~\ref{eqn:Efder1}, where $\omega_p^2=ne^2/m_b\epsilon_0$, we find an expression for E$_F$ in [eV] as a function of physical constants, $\omega_p^2$ in [rad/s] and $n$ in [m$^{-3}$]:
%
%
\begin{equation}\label{eqn:Efder2}
E_F=\frac{\hbar^2 \omega_p^2\epsilon_0}{2e^3 n^{1/3}}\left(\frac{6\pi^2}{g_s g_v}\right)^{2/3}
\end{equation}
%
%
Eqn.~\ref{eqn:Efder2} yields a value for E$_F$ ($\approx\mu$) of 126 meV at 5 K, which is consistent with our value of $\mu$ of $114\pm22$ meV obtained from the direct gap transitions (see section.~\ref{sec:PSSinterband}), and further confirms the metallic character of naturally doped n-type Pb$_{0.77}$Sn$_{0.23}$Se. As explained in Section~\ref{sec:PSSss}, this further confirms why surface states are not observed in the Reflectance spectra. 

\subsection{Power Factor}
\label{sec:Powerfactor}
\begin{figure}
\includegraphics[scale=0.31]{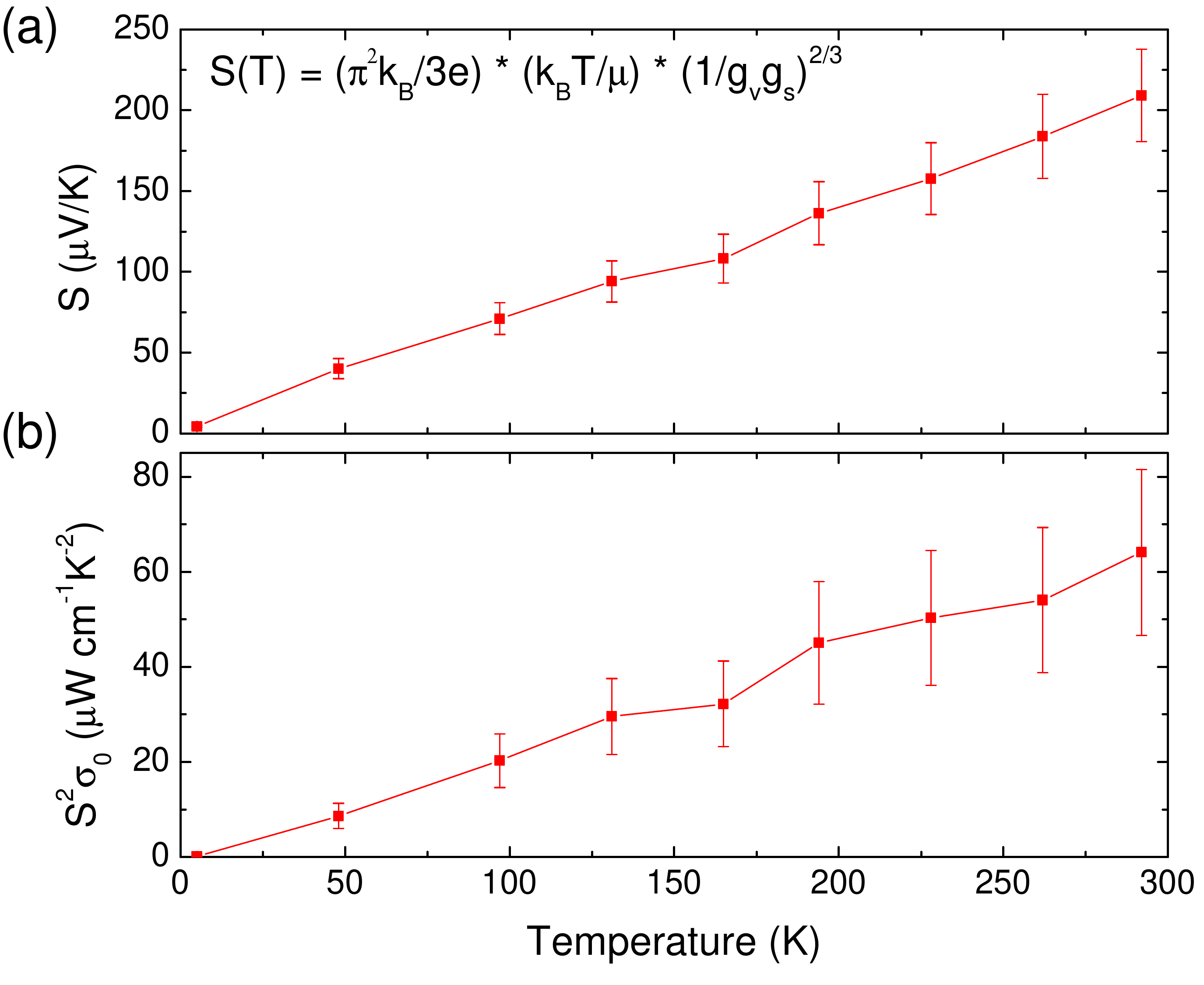}
\caption[The Seebeck coefficient and power factor of Pb$_{0.77}$Sn$_{0.23}$Se]{\label{fig:PSSFig5} (a) Seebeck coefficient in the parabolic band assumption, where all changes in $E_{dir}$ are attributed to changes in $E_F$. (b) The Power Factor indicates minimal (if any) sensitivity to the band inversion at $160\pm15$ K, as a result of the high chemical potential.
}
\end{figure}
Finally, we relate our obtained values for $\sigma_{o}$ and $\mu$ to the Seebeck coefficient $S(T)$ and the power factor $S^2\sigma_o(T)$. In the parabolic band approximation, the Seebeck coefficient is given by\cite{2008NatMa...7..105S} $S=\tfrac{8\pi^2k_B^2}{3eh^2}m_bT(\tfrac{\pi}{3n})^{2/3}$ (where $m_b$ is the conduction of valance band mass), which can be expressed as a function of $E_F$ (using eqn.~\ref{eqn:Efder1}) as
%
%
\begin{equation}\label{eqn:Seebeck}
S(T)=\left(\frac{\pi^2k_B}{3e}\right)\left(\frac{k_BT}{E_F}\right)\left(\frac{1}{g_sg_v}\right)^{2/3}
\end{equation}
%
%
Since $\mu$ is located high in the conduction band, the direct gap transitions are likely only minimally sensitive to changes in the true band gap $E_g$ (and thus $m_c$) as a function of temperature. This is illustrated in the left and right panel in Fig.~\ref{fig:PSSFig2}a, showing a similar band dispersion for high $\mu$ despite a dramatic change in the band gap. Hence, we can estimate the maximum change in $S(T)$ by attributing the temperature dependence of $E_{dir}(T)$ (see eqn.~\ref{eqn:directgap}) entirely to changes in $\mu(T)$. The result of this estimation, after substituting $\mu(T)$ for $E_F$ in eqn.~\ref{eqn:Seebeck}, is plotted in Fig.~\ref{fig:PSSFig5}a, while Fig.~\ref{fig:PSSFig5}b shows the associated power factor $S^2\sigma_0$. It is clear that neither $S$ nor $S^2\sigma_0$ show distinct signatures of the band inversion at $160\pm15$ K. These results suggests that there is room for optimization of the figure of merit $ZT$ in Pb$_{1-x}$Sn$_{x}$Se compounds where $\mu$ is located deep in the conduction band. Further doping to alter the lattice and suppress phonon contributions will likely only affect the electronic response minimally, as illustrated by its insensitivity to the band inversion and the unaffected power factor.  

\

\section{Conclusion}

In conclusion, we have performed broadband optical spectroscopy experiments on as-grown n-type Pb$_{0.77}$Sn$_{0.23}$Se and extracted its optical constants over a temperature range between 5 K and 293 K. By studying the plasma frequency and band gap transitions,  we find clear signatures of the temperature dependent band inversion at $160\pm15$ K, despite the chemical potential being far in the conduction band. We note that since temperature is an easily controllable variable in optical spectroscopy, in principle this could be used to optically determine band inversion temperatures of conductive samples with extremely high precision.
While this band inversion is also associated with a topological phase transition,\cite{2011PhRvL.106j6802F,2012NatMa..11.1023D,Wojek:2013hx,2013arXiv1306.0043G,2013arXiv1305.2823O,Liang:2013tk,Xi:2014uz} we do not see any signatures of surface states, and thus don't comment on the topology of the band structure on either side of the band inversion. This is an expected result, however, as optical modelling suggests that the highly conductive bulk carriers dominate the properties of as-grown Pb$_{0.77}$Sn$_{0.23}$Se. 

By performing extended Drude analysis, we accurately determine the temperature dependent scattering rate, and find that it follows $T^{3/2}$ behaviour (associated with electron-acoustic phonon interactions) typical for good thermoelectrics,\cite{Pei:2011kx} superimposed on a constant impurity scattering background with $1/\tau=6.8\pm0.3$ meV. The frequency dependence of the scattering rate shows no evidence of different scattering channels below and above the band inversion, but data over a greater energy and temperature range are required for more conclusive results. Moreover, neither the DC resistivity, Seebeck coefficient, or the power factor show any evidence of the band inversion. This result confirms earlier transport studies on similar compounds, \cite{Shukla:2007tt,Ekuma:2012gl,Gibbs:2013hq,1990JPCM....2.2935B,1967PhRv..157..608S,Zhao:2014du,Butler:1966vb} and emphasizes the utility of optical spectroscopy in the study of thermoelectrics and/or band inversions.

For application of Pb$_{0.77}$Sn$_{0.23}$Se as either a topological crystalline insulator or thermoelectric, the high electrical conductivity of bulk states is a major prohibitive factor. Moreover, some of the fundamental physics governing the electronic, thermal, and topological properties are difficult to probe due to the high chemical potential in as-grown Pb$_{0.77}$Sn$_{0.23}$Se. Hence, future studies on samples in the Pb$_{1-x}$Sn$_{x}$Se series, in which the carrier density has been reduced, could lead to new insights into the interplay of phonons, thermal and electronic bulk lifetimes, and the optical and electronic properties of inversion symmetry protected topological surface states. 

A reduction of the chemical potential would also enhance the sensitivity of the Seebeck coefficient to the band inversion, which could potentially result in a dramatic enhancement of the figure of merit $ZT$ around the band inversion. Moreover, alternative stoichiometries also offer an interesting avenue, as previous measurements suggest that the band inversion in Pb$_{1-x}$Sn$_{x}$Se could be pushed to higher temperatures -relevant to many thermoelectric applications- by increasing the Sn content.\cite{2012NatMa..11.1023D,1967PhRv..157..608S,1971PhRvB...3.4299D} Hence, by optimizing Se doping, an ideal Pb$_{1-x}$Sn$_{x}$Se stoichiometry can be determined, in which $ZT$ peaks around a desired temperature range. Indeed, such doping studies would also allow for the study of the high temperature interplay between the topologically nontrivial phase and thermoelectric efficiency, which is a direction currently unexplored.

With both thermoelectrics and topological (crystalline) insulators offering numerous appealing applications in industry, while presenting nontrivial physical characterization and optimization challenges, a better understanding of fundamental physics governing the figure of merit $ZT$ and the isolation of surface conductance are both of practical and fundamental interest. The Pb$_{1-x}$Sn$_{x}$Se and Pb$_{1-x}$Sn$_{x}$Te series offer many exciting opportunities in this pursuit, particularly through optical investigations, by combining several novel physical phenomena into a single temperature tuneable system.
\

\section{Acknowledgements}
We acknowledge Young-June Kim for use of an ellipsometer, as well as Luke Sandilands for very helpful discussions. This work has been partially funded by the Ontario Research Fund, the Natural Sciences and Engineering Research council of Canada, Canada Foundation for Innovation, and the Prins Bernhard Cultuurfonds. The crystal growth at Princeton University was supported by grant N6601-11-1-4110.

We also acknowledge the recent optical and Hall-effect study of Pb$_{0.77}$Sn$_{0.23}$Se by Anand {\it et al.},\cite{Anand:2014vn} which was performed simultaneously, though independently. This study explores the temperature-driven band inversion of p-type Pb$_{0.77}$Sn$_{0.23}$Se, which is found to occur around 100 K.

%

\newpage
\appendix

\section{Expected optical signatures of TCI surface states in Pb$_{0.77}$Sn$_{0.23}$Se}
%
\begin{figure}
\includegraphics[scale=0.31]{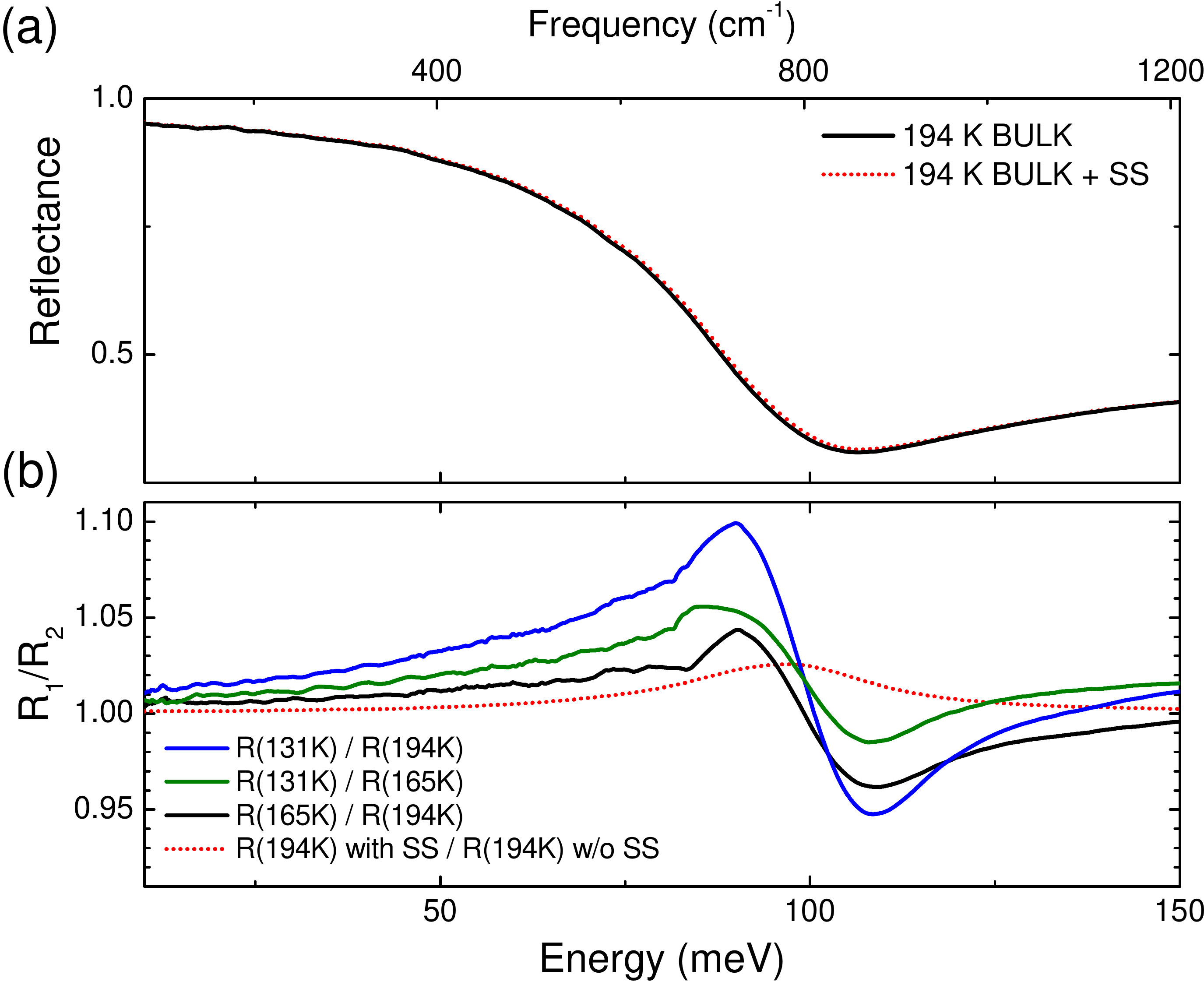}
\caption{\label{fig:PSSFig2p5} (a) Reflectance of Pb$_{0.77}$Sn$_{0.23}$Se at 194 K, and a model of the 194 K Reflectance that included a surface state layer with $\omega_{p,2D}=0.34$ meV cm$^{1/2}$. (b) Ratios of the temperature dependent reflectance above (black), below (green), and across (blue) the topological phase transition. The dotted red line is the ratio of reflectance spectra from (a), illustrating the much larger effect of temperature on $R(\omega)$ compared to the introduction of surface states below the topological phase transition.}
\end{figure}
%
To ensure we did not overlook surface state signatures in our reflectance data, we use the Boltzmann transport equation to estimate the expected effect of surface states on the measured reflectance. The 2D surface state plasma frequency can be approximated by $\omega_{p,2D}^2=(e^2/h)2\pi k_F v_F g_s g_v$, where $v_F$ and $k_F$ are the Fermi velocity and wavevector, and $g_s$ and $g_v$ are the spin and valley degeneracies.\cite{BTSpaper,Chapler,2007PNAS..10418392A} Ignoring the parabolic correction for the Dirac states, we can use $E_F=\hbar k_F v_F$, and express $\hbar\omega_{p,2D}$ in a convenient form with all SI units
%
%
\begin{equation}\label{eqn:PSS2DWp}
\hbar\omega_{p,2D}=c\sqrt{eE_F g_s g_v10^{-5}}
\end{equation}
%
%
where $c$ is the speed of light and $e$ is the elementary charge. From previous ARPES measurements, we can estimate a surface Fermi level $E_F^{ss}$ around 125 meV.\cite{2012NatMa..11.1023D} At the risk of overestimating the surface contribution, we can account for sample-to-sample variations by considering an even higher surface state Fermi level of 200 meV. We also note that while $g_s=1$ and $g_v=2$ near the Dirac point, additional Dirac states with smaller wavevectors intersect the Fermi surface when the Fermi level crosses the bulk conduction band, thus doubling the surface bands, albeit with different wavevectors.\cite{2012NatMa..11.1023D,Liang:2013tk,2013arXiv1305.2823O,Liu:2013jh} For convenience, we account for these states by setting $g_v=4$, resulting in a overestimated 2D plasma frequency of 0.34 meV cm$^{1/2}$. 

Fig.~\ref{fig:PSSFig2p5}a shows the reflectance of Pb$_{0.77}$Sn$_{0.23}$Se at 192 K (the topologically trivial regime) as well as a model in which a surface state layer with $\hbar\omega_{p,2D}=0.34$ meV cm$^{1/2}$ was added to the 192 K Reflectance.  To emphasize the minimal difference, Fig.~\ref{fig:PSSFig2p5}b shows the ratio of both reflectance spectra (red dotted line). The same plot also shows the ratio of reflectance measurements at different temperatures above (black), below (green), and across (blue) the topological phase transition. It is clear that the sharpening bulk plasma edge has a far greater effect on the reflectance, than the surface states. Hence, discriminating surface state signatures are not within our measurement capabilities for samples with such high bulk carrier densities.

\end{document}